# Electronically Steered Nyquist Metasurface Antenna

Michael Boyarsky*, Timothy Sleasman, Mohammadreza F. Imani, Jonah N. Gollub, & David R. Smith

Center for Metamaterials and Integrated Plasmonics, Department of Electrical and Computer Engineering, Duke University, Durham, NC 27708

*michaelboyarsky@gmail.com

Mobile devices, climate science, and autonomous vehicles all require advanced microwave antennas for imaging, radar, and wireless communications. The cost, size, and power consumption of existing technology, however, has hindered the ubiquity of electronically steered systems. Here, we propose a metasurface antenna design paradigm that enables electronic beamsteering from a passive lightweight circuit board with varactor-tuned elements. Distinct from previous metasurfaces (which require dense element spacing), the proposed design uses Nyquist spatial sampling of half a wavelength. We detail the design of this *Nyquist metasurface antenna* and experimentally validate its ability to electronically steer in two directions. Nyquist metasurface antennas can realize high performance without costly and power hungry phase shifters, making them a compelling technology for future antenna hardware.

# ARTICLE

Electronic beamsteering is an essential capability for antennas used in Earth observation, radar, and communications [1-6]. A common means of forming a desired radiation pattern is to specify the phase and amplitude of the field over an aperture. Fourier optics then provides the quantitative connection between the spatial distribution of the aperture field and the angular distribution of the far-field. When the phase and magnitude of the aperture fields can be specified without constraint, the possible far-field radiation spatial patterns are nearly boundless, subject only to diffraction limits [4-5].

Aperture antennas that generate and steer beams or other tailored patterns inherently make use of this Fourier relationship. In practice, it is the phase of the aperture fields that has much more of an influence on the far-field waveform, leading naturally to the concept of the phased-array antenna [4-5]. A phased array antenna achieves its beam steering capabilities through active phase shifters—devices that require external power—positioned at every radiating node across a defined aperture [4-5,8]. Exactly controlling the phase of each radiating node in this way provides excellent electronic beamforming capabilities.

The number of radiating nodes is typically set by the Nyquist theorem, which states that a signal needs to be sampled at a rate twice the highest frequency component present. For aperture antennas, this requirement translates to spatial sampling of half of the operational wavelength across the aperture (depending on the desired steering limits). In antenna systems where both the phase and magnitude are controlled, such as in more advanced electronically scanned antennas (ESAs), amplifiers, circulators, and other components are often present at each node, resulting in high performance, but at the cost of considerable system complexity, cost, and power draw [4-5,8].

The system complexity of ESAs has led to the development of many alternatives which exhibit reduced capabilities. Reflector dish systems can steer a beam with motors, but this operation suffers from slow switching speeds and limited beam tailoring capabilities [9]. Leaky-wave antennas can form a beam with a series of irises etched into a waveguide, but struggle to form arbitrary radiation patterns independent of frequency [10-17].

Recently, metamaterials and metasurfaces have gained attention as a new type of electromagnetic device. While initially considered as artificial materials described by effective material parameters, metamaterials have since proven advantageous as the basis for a wide range of electromagnetic products [18-21]. Specifically, metasurface apertures have been developed as a type of holographic antenna, using metamaterial elements to form a hologram excited by a feed wave acting as a reference. Recently, a waveguide-fed metasurface antenna was created for satellite communication which exhibited high performance beamforming capabilities [21]. Though existing demonstrations have shown the great promise of metasurface antennas, limitations in efficiency, modeling, and switching speed have limited their deployment.

Waveguide-fed metasurfaces use a waveguide mode to excite metamaterial radiators etched into one of the conducting walls. The incident field drives a resonance in each element, selectively leaking energy out of the waveguide. The overall radiation pattern of the aperture is then the superposition of the radiation from each element [22-25]. Externally tunable components, such as liquid crystal or diodes, can change the response of each element independently by shifting the resonance. Tuning each element allows the overall aperture's response to be dynamically reconfigured, enabling the rapid creation of arbitrary radiation patterns, including steerable, directive beams [20-27].

Although metasurfaces have demonstrated electronic beamforming, the question remains as to how their performance compares with a true phased array. While an active phase shifter can tune the phase over a range of 0-360°, a passive, resonant metamaterial element can, at best, tune across a 0-180° range [4-5,23]. Further, the magnitude and phase response of a metamaterial element are linked through its resonance. Thus, the phase and magnitude of a passive, radiating element cannot be controlled independently [23,27].

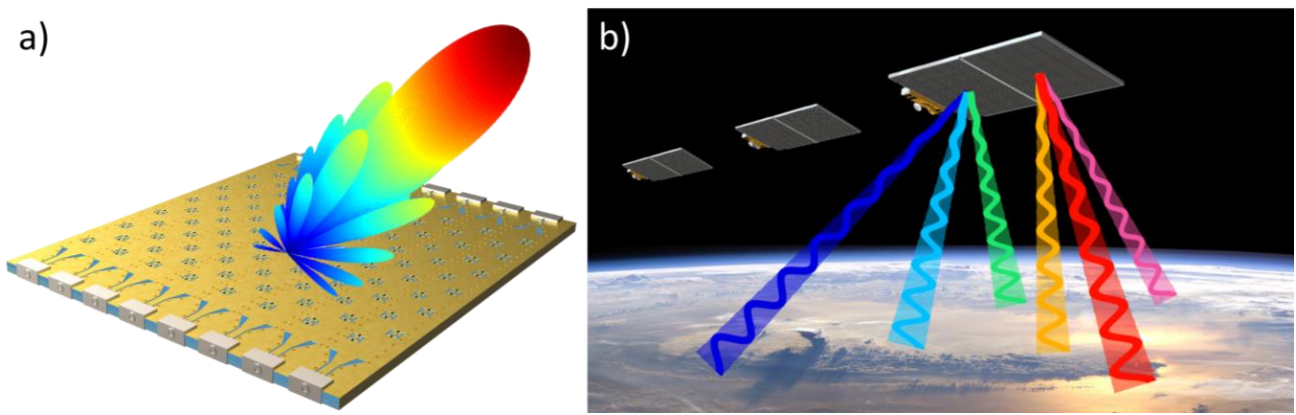

**Figure 1 | Nyquist metasurface antenna. a)** shows a Nyquist metasurface antenna forming a steered beam. Nyquist metasurface antennas offer high performance from a low cost and thin platform, enabling the construction of flat satellites. **B)** shows a multi-satellite system using Nyquist metasurface antennas for satellite-to-ground communications.

Center for Metamamaterials and Integrated Plasmonics, Department of Electrical and Computer Engineering
Duke University, Durham, North Carolina 27708, USA

Despite this constrained control, waveguide-fed metasurface antenna architectures have demonstrated high quality beamforming by compensating for the reduced phase range by densely sampling the aperture (typically on the order of one-sixth or less of the operating wavelength) and leveraging the phase advance of the guided wave [23, 23-27].

In this paper, we describe the design of a metasurface antenna which relies on two key concepts: feed phase diversity and varactors diodes. Feed phase diversity involves offsetting the initial wave in each waveguide comprising a metasurface antenna array; this strategy allows for the suppression of grating lobes without dense element spacing [27]. Meanwhile, varactor diodes enable continuous phase tuning from metamaterial elements to provide improved element tuning compared to PIN diodes [28]. This design paradigm ultimately allows for metamaterial element placement at roughly half the wavelength of operation. Given that such sampling corresponds to the Nyquist limit, we describe this device as a *Nyquist metasurface antenna*.

Here, we demonstrate a Nyquist metasurface antenna, fabricated using standard printed circuit board (PCB) manufacturing. We show electronic beamforming in two (angular) dimensions while operating at 10 GHz. We show the Nyquist metasurface antenna has excellent performance, without incurring the high cost, power consumption, and complexity of typical phased array antennas. Furthermore, the proposed Nyquist metasurface antenna can be readily redesigned to operate at higher or lower frequency bands and scaled to very large apertures. An illustration of the antenna is shown in Figure 1a. Figure 1b shows the potential to leverage the hardware characteristics and performance associated with Nyquist metasurface antennas to deploy a constellation of flat satellites for communication.

## Methods

**Metasurface antenna operation and challenges.** To arrive at the design for the Nyquist metasurface antenna, we addressed three major challenges associated with metasurface antennas: modeling the antenna, suppressing grating lobes, and minimizing the coupling strength of each element. In this section, we describe these challenges, which motivated our design, before briefly outlining our solutions. Further detail on the antenna design is provided in the next section. By applying a cohesive design approach that addressed these issues simultaneously, we were able to realize a new design paradigm for high performance metasurface antennas.

First, ESAs often comprise electrically large structures with subwavelength features, rendering them difficult to model with full-wave electromagnetic solvers, since the simulation domain tends to be extreme. A common modeling approach is to simulate the radiation pattern of a single element in isolation. A composite antenna array can then be predicted using array factor calculations to predict radiation patterns. These method works well if the feed wave at every node is known and each element behaves the same.

In metasurface antennas, an approximate model is further necessary due to the small feature size of metamaterial radiators (often $<\lambda/10$). Additional consideration must be given to the guided wave in metasurfaces. As the guided wave sequentially excites metamaterial elements along a waveguide, they scatter a portion of its energy, leading to gradual decay. In our model, each metamaterial element is modeled as frequency-dependent point dipole whose response can be measured from full-wave simulation (of one element) or experimental characterization. Once the response of a single element has been determined, an overall antenna can be modeled as a series of dipoles with array factor calculations. This approach enables the rapid simulation of different antenna layouts, tunings, and frequencies [29-30].

Second, metasurface antennas are especially susceptible to unwanted grating lobes due to the lack of complete control over phase and magnitude. In traditional antennas, grating lobes can arise when attempting to steer beyond what the element spacing will allow. When sampled at the Nyquist limit, there is no upper bound on steering, but sparser antennas can face this challenge. Metasurfaces, however, face grating lobe problems due to their incomplete phase control – not their physical spacing. These metasurface-specific grating lobes arise because phase-tuning metamaterial elements for beamforming creates an incidental, oscillating magnitude profile. Since many elements are tuned to essentially non-radiating states, the effective element spacing becomes coarse, leading to metasurface-specific grating lobes, independent of physical element separation [27].

While previous metasurface designs have suppressed metasurface grating lobes using a combination of high dielectric loading (to increase the effective waveguide index) and dense element spacing, here instead we seek a design that allows a sparser sampling (close to half of a wavelength). We rely on the feed structure to provide phase diversity at the excitation of each waveguide in our array. Such feed phase diversity suppresses grating lobes in 2D metasurface arrays regardless of whether elements are spaced closer than the Nyquist limit [27]. With grating lobes suppressed by the feed structure, the elements can be positioned with spacing at or near the Nyquist limit, even if hollow waveguides are used.

Third, as with leaky-wave antennas, the incident waveguide mode loses energy as it propagates through the structure and excites the radiating metamaterial elements. If the elements are strongly coupled to the waveguide mode, the waveguide mode attenuates quickly and much of the aperture no longer radiates significantly, resulting in a reduced effective aperture size. To avoid this effect, the coupling of the metamaterial elements can be carefully reduced by offsetting them from the center of the waveguide (and further with tuning algorithms). Offsetting the elements from the center and alternating sides of the waveguide reduces the coupling of the elements to the waveguide mode, allowing energy to reach elements far from the feed. Additionally, this approach increases the distance between any two elements, reducing inter-element coupling, increasing the fidelity of the modeling approach used in this work [23,25,27,30].

**Antenna Subsystems.** The composite 2D Nyquist metasurface antenna consists of an array of 1D waveguides, which each excite a linear array of metamaterial resonators.

Center for Metamamaterials and Integrated Plasmonics, Department of Electrical and Computer Engineering
Duke University, Durham, North Carolina 27708, USA

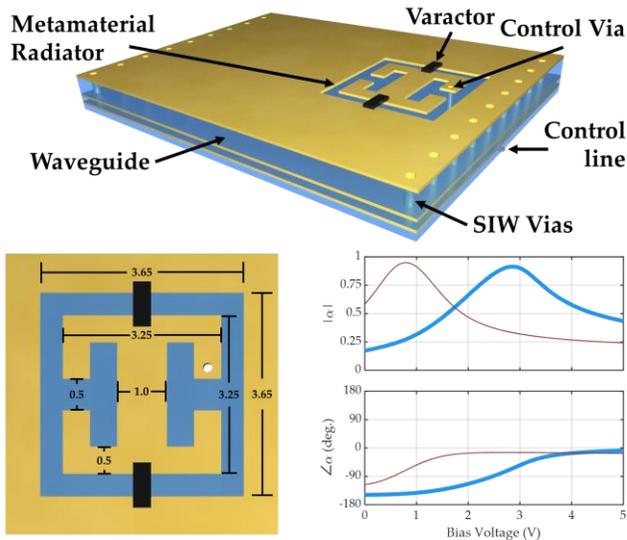

Figure 2 | Varactor-tuned metamaterial element. The bottom left figure shows the dimensions of the metamaterial radiators (in mm). The bottom right figure shows the magnitude and phase of the experimentally characterized element response, measured at 9 GHz (red) and 10 GHz (blue), as the tuning sweeps from 0 to 5V.

The waveguides are substrate integrated waveguides (SIWs), which use a pair of via fences and metal layers to form a rectangular waveguide within a PCB. The tunable element design, waveguide architecture, feed structure, and control system must all be designed in an integrated fashion.

The metamaterial element design (described in more detail in [28]) is a complementary electric-inductive-capacitive (cELC) resonator with outer dimensions of 3.65 mm by 3.65 mm, as shown in Figure 2. The cELC is used because it behaves electromagnetically as a polarizable magnetic dipole with a resonant polarizability, which can be electronically tuned. Varactors placed across the capacitive gaps between the metamaterial and the surrounding waveguide's upper conductor provide a means of tuning the element's capacitance, thereby tuning its resonance. The main considerations for choosing the varactor are package size and self-resonant frequency. Given the relatively high frequency range we are targeting, it is important that the self-resonant frequency be significantly higher than the operating frequencies so that the varactor does not add additional inductance or resistance to the circuit. MACOM varactors (MAVR011020) were found to satisfy the requirements and selected for this design. A bias circuit is integrated into the element design, with a control via extending from the center of the cELC through the SIW core and through the bottom conductor of the waveguide to a layer used exclusively for biasing circuitry. Note that the control via is located near the edge of the SIW to reduce its impact on the guided wave.

Applying voltage between 0 – 5 V changes the overall capacitance of the cELC and shifts the resonance of the element from 8.5 GHz to 10.7 GHz. At 10 GHz, the primary operating frequency of this antenna, this tunability equates to 150° of phase tuning (the theoretical maximum phase tuning of a Lorentzian resonator is 180°) and a magnitude ratio of 4.5:1. The varactor-tuned metamaterial element is illustrated in Figure 2, which also shows the voltage-tuned polarizability, experimentally characterized at 10 GHz.

Substrate integrated waveguides were chosen for the metasurface antenna design as they are easily fabricated using commercial, multilayer PCB technology (see Figure 3c). SIWs behave as rectangular waveguides with fields well-confined and described by well-known analytical expressions [32]. Rectangular waveguide modes are particularly helpful in the context of a Nyquist metasurface antenna as most of the energy in the waveguide mode is concentrated towards the center. As a result, this waveguide structure allows for metamaterial elements to be offset from the center of the waveguide to decrease coupling in order to allow sufficient energy to pass to subsequent elements.

The Nyquist metasurface antenna layout consists of tiling the aperture area with adjacent SIWs; in this way, a 2D aperture antenna comprises an array of eight 1D SIWs. Each SIW is 14 mm, with the SIWs separated by 15 mm to provide space for via fences (see Figure 4). To launch a wave into an SIW, we use end launch connectors (shown in Figure 3a). The end launch connector excites a grounded coplanar waveguide (CPW) mode, which subsequently feeds the SIW through a CPW-to-SIW transition, optimized in CST Microwave Studio following [33], and shown in Figure 3b. The components of the waveguide structure are detailed graphically in Figure 3. A 50Ω terminator at the end of each waveguide minimizes reflection by absorbing the remaining energy.

One of the challenges mentioned above is suppressing metasurface-specific grating lobes [27]. Metasurface antennas are tuned with passive components to avoid active phase shifters, sacrificing complete phase control for a continuous but limited range of phase tuning (less than half that of a phase shifter). The impact of this limitation can be understood when considering the formation of a beam steered

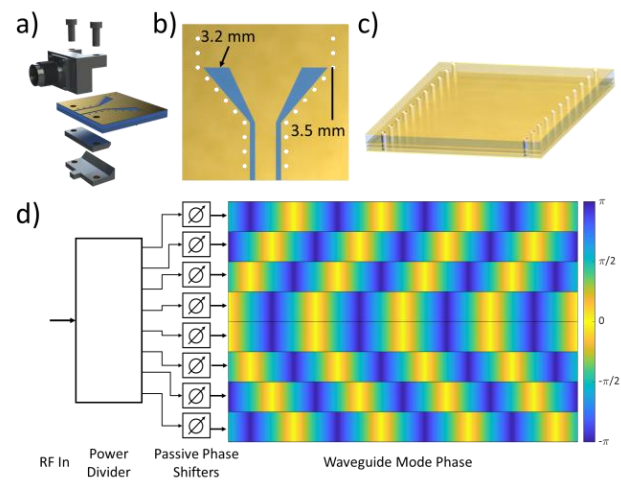

Figure 3 | Waveguide structure underlying the Nyquist metasurface antenna. a) shows an end launch connector which excites a CPW. b) shows the transition from CPW to the SIW shown in c). d) shows the RF feed network which uses an eight-way power divider and passive phase shifters to apply the requisite feed phase diversity.

Center for Metamamaterials and Integrated Plasmonics, Department of Electrical and Computer Engineering
Duke University, Durham, North Carolina 27708, USA

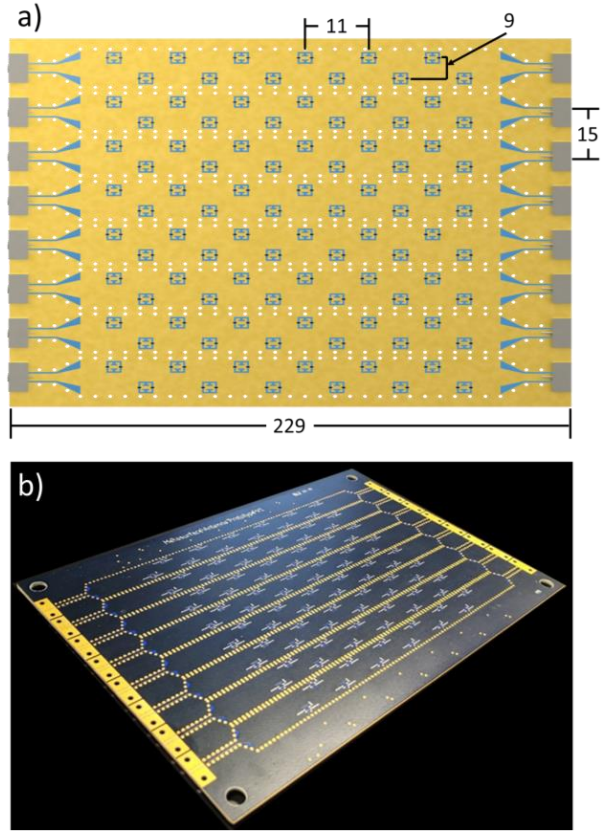

**Figure 4 |** Nyquist metasurface antenna layout, a), with dimensions in mm. b) shows a picture of the antenna.

to some angle, which requires an aperture field distribution whose phase profile linearly increases as a function of position. For metasurface antennas, simply attempting to match this phase distribution leads to approximately half of the elements being unused; these elements are set to non-radiating states since the targeted phase values lie outside the available phase space, which locks these element to the voltage-tuned limit (and has low magnitude). This phenomenon leads to an oscillating magnitude profile due to the repeating collections of elements tuned to non-radiating states. In a recent work, it has been found that if multiple waveguides are used, each excited with a distinct phase, the grating lobes can be cancelled out [27]. The combination of this *feed diversity* with optimized tuning strategies can lead to a 2D metasurface antenna that fully suppresses metasurface grating lobes without relying on dense element spacing.

The necessary phase shift per waveguide could be achieved using an integrated corporate feed layer that would act as a power divider. Rather than working through this more complicated feed design, we chose for expediency to use an eight-way power divider connected to a series of passive, mechanically set phase shifters to create the requisite phase diversity. From top to bottom, the phase shifters provide a feed phase of [270°, 180°, 90°, 0°, 0°, 90°, 180°, 270°], respectively, resulting in the feed waves shown in Figure 3d. The phase shifters have been mechanically set for operation at 10 GHz. A more elegant waveguide feed structure could operate over a larger bandwidth, but such a structure is beyond the scope of this work.

**Modeling and tuning.** An accurate model of a metasurface antenna is necessary to determine the tuning of each element needed to form desired radiation patterns such as steerable, directive beams. For electrically large antennas, typical simulators that numerically solve Maxwell's equations require extremely large numbers of unknowns resulting in prohibitively long simulation times and memory storage requirements. As a means of dealing with the multiscale modeling problem, we abstract each metamaterial element as a frequency-dependent, infinitesimal, polarizable dipole, as mentioned above [29-30]. The radiated fields can then be quickly and easily determined by summing the radiated fields from each of the effective dipoles. During the design process, the polarizability of one element was determined from a full-wave simulation of one element, requiring a minimal simulation domain.

The dipole moment representing each metamaterial element (η) can be calculated as the product of the incident magnetic field (H) and the element's polarizability (α) [30].

$$\eta = H\alpha \qquad (1)$$

Metamaterial elements are resonant structures. If the element is suitably smaller than the operational wavelength, it can be modeled as a polarizable dipole as a function of geometry and material parameters [23]. The magnetic polarizability of a cELC element follows a Lorentzian response as a function of the excitation frequency (ω), resonant frequency ($\omega_0$), coupling factor (F), and damping (γ).

$$\alpha = \frac{F\omega^2}{\omega_0^2 - \omega^2 + j\omega\gamma} \qquad (2)$$

These parameters are difficult to estimate analytically but can be readily obtained from full-wave simulations of single metamaterial elements [25,30]. Those useful for prediction and design, experimental characterization must ultimately be used to determine α accurately for a given metasurface structure, since fabrication variances and varactor properties lead to unavoidable uncertainties for key parameters.

After fabrication, experimental characterization determined the exact value of α for the metamaterial element. A separate PCB was fabricated alongside the metasurface antenna to facilitate waveguide and element characterization. This characterization board, which is detailed further in [22], includes thru-reflect-line (TRL) characterization channels as well as a channel with a single element. The TRL channels provide a means of de-embedding an individual element's scattering parameters to isolate them from the surrounding waveguide structure [34,35]. The scattering parameters can then be used to calculate the element's polarizability as a function of frequency and tuning voltage [25,36-37].

$$\alpha = -\frac{jab}{\beta}(1 + S_{11} - S_{21}) \qquad (3)$$

Here, a and b are the horizontal and vertical dimensions of the waveguide, respectively, and β is the wavenumber of the SIW

# ARTICLE

(determined experimentally). Figure 2 shows the element response at 10 GHz, showing the tunable α. Polarizability extraction in this way is further detailed in [25,28].

After measuring or modeling α, the guided wave, H, must be modeled. The waveguide mode can be readily calculated analytically as a function of the geometry and material properties of the waveguide. To validate the analytic model of H, the characterization board described above was also used to experimentally measure β of an empty waveguide.

When a waveguide is populated with several metamaterial elements, scattering from each element decays the guided wave's energy and applies a phase shift. To account for the impact of each element's scattering on the guided wave, complex $S_{21}$ is extracted along with α during the element characterization. The field at element $n$, $H_n$, can then be determined approximately as a function of each element's tuning and position.

$$H_n = H_0 e^{-j\beta x_n} \prod_{n<N} S_{21,n} \quad (4)$$

Here, $N$ is the total number of elements (along one waveguide), $H_0$ is the initial magnetic field, and $x_n$ is the position of the $n$th element along the waveguide. This equation for $H_n$ can be treated as a perturbative model calculated sequentially to accurately include the impact of each element's tuning state on the guided wave. Note that underlying this method, as evidenced by the linearity of the model, is the assumption that the elements are non-interacting. Both the antenna layout and tuning strategies are designed towards the goal of minimizing inter-element coupling to strengthen the assumption of linearity.

As shown in Eq. 2, a metamaterial element's magnitude and phase response are linked through their resonance. Tuning the phase of each element to form a desired aperture field incidentally applies an unwanted, oscillating magnitude profile, which leads to grating lobes if not properly considered [27]. Further, the maximum phase shift available to a metamaterial element is only 180°. To account for these restrictions, specific tuning strategies must be applied to the metasurface antenna to realize optimal performance.

To form a beam in the far-field from a 2D aperture, the complex antenna weights must exhibit this phase profile.

$$\eta_{desired,nm} = e^{jk(x_n \sin\theta_0 \cos\varphi_0 + y_m \sin\theta_0 \sin\varphi_0)} \quad (5)$$

Here, $k$ is the free space wavenumber, $\theta_0$ and $\varphi_0$ are the angles to which the beam is steered, and $x_n$ and $y_m$ are the position [4-5,23]. A magnitude taper can also be applied in order to change the side lobe levels and other antenna metrics, but such considerations are beyond the scope of this investigation. Equation 5 is the starting point to determine the tuning state of a metasurface antenna, in which we have control over the polarizability of each element. The product of the polarizability and the incident magnetic field equate to the dipole moment (Eq. 1), which functions in exactly the same manner as the antenna weights in an array factor calculation. For a given metamaterial element to have the value $\eta_{desired,nm}$, the element's tunable polarizability must be chosen so as to counteract the phase advance of the guided wave and apply the phase prescribed by Eq. 5. In a traditional phased array antenna, this process would be done by simply mapping the phase of each element to that of Eq. 5. Here, we opt for a metasurface-specific tuning strategy—Euclidean modulation—which maps each element's polarizability by minimizing the Euclidean norm between the desired and available values of $\alpha_{nm}$ (rather than minimizing the phase difference). Using this mapping strategy leads to a compromise between the phase and magnitude response of a metamaterial element, resulting in highly directive beamforming while avoiding unnecessary efficiency loss.

In a metasurface antenna, as the guided wave traverses the waveguide, energy is gradually leaked into free space, creating a natural exponential magnitude taper. The decay rate of the taper is related to the tuning state of each element, the average coupling strength, the dielectric loss, and the element spacing. This decay can limit performance in two major ways. If the decay is too fast, a portion of the aperture will be unused, limiting aperture size and thus directivity. If the decay is too slow, unused energy will be terminated at the end of the waveguide, reducing efficiency. To balance these factors, the decay of the guided wave must be deliberately controlled with tuning and geometry. A balance between these concerns can be met by augmenting Euclidean modulation with a scale factor. To use scaled Euclidean modulation, $\alpha_{desired,nm}$ is divided by a scale factor ($A$) as $\alpha_{desired,nm}/A$ before minimizing applying Euclidean modulation.

$$\min \left| \frac{\alpha_{desired}}{A} - \alpha_{available} \right| \quad (6)$$

The value of $A$ used in this work was obtained empirically.

**Nyquist metasurface antenna design.** The final metasurface antenna design comprises eight adjacent waveguides that cover a 2D area. Each waveguide includes an integrated waveguide transition show in Figure 3. A coaxial-fed end launch connector excites a wave into a grounded coplanar waveguide, which then transitions to a SIW. Each waveguide contains 12 individually tunable metamaterial elements, spaced 11 mm apart along the waveguides. As indicated in Figure 2, each element is located 4.75 mm away from the center of the waveguide to reduce coupling with the guided wave. Additionally, elements alternate sides down the waveguide to increase the distance between elements, reducing the inter-element coupling. The overall layout is illustrated in Figure 4.

The antenna is fabricated on a four-layer PCB using Rogers 4003C (ε=3.38, δ=0.0027). The top two layers contain the waveguides and metamaterial elements, while the bottom two layers contain control circuitry and components. The elements are controlled using 8-bit, 8 channel digital to analog converters (DACs), which provide an independent bias for each element from 0 to 5V. A PC running Matlab was interfaced with an Arduino microcontroller to control the antenna. Radial stubs are connected to each control line to decouple the DC and RF signals. The antenna was designed to operate over a bandwidth of 9.6 to 10 GHz.

The overall Nyquist metasurface antenna uses 96 elements to cover a radiating area of 12 cm by 12 cm. The sparse layout has not been presented in previous metasurface antenna designs due to concerns with grating lobes. But by combining feed diversity (for grating lobe suppression) with continuous phase tunable metamaterial elements (using varactors), such a Nyquist metasurface antenna can avoid the high dielectrics and dense element spacing associated with prior waveguide-fed metasurface antenna designs. The sparse layout of weakly coupled elements has the additional advantage that the analytical dipole model does not need to consider interactions among elements. Though such modeling can be performed using coupled dipoles, simulation and optimization are much more straightforward if each element can be treated as an independent dipole.

**Results**

The main design goal for the metasurface antenna described in this work was to demonstrate the generation of a directive beam that could be steered in azimuth and elevation over a defined bandwidth. Efficiency, sidelobe level, and all other metrics were secondary considerations, though they could readily be addressed in future designs. After the characterization board was used to measure a single element and an empty waveguide, those parameters were incorporated into the mathematical model of the antenna. For the metamaterial element, the effective polarizability was extracted as a function of frequency and tuning voltage. For the feed structure and empty waveguides, this meant that the experimentally characterized waveguide mode was used in place of the analytically modeled $H$ (including the phase diversity present within $H_o$). Scaled Euclidean modulation optimization (Eq. 6) was then applied to determine the tuning state for each element as a function of steered angle. After the determining each element's tuning state, the appropriate bias voltage was then applied to each element in the antenna.

To characterize the radiation pattern of the antenna, near-field scan measurements were taken at an independent facility with an anechoic chamber (Wireless Research Center, Wake Forest, NC). The near-field measurements were then propagated to the far-field to determine the antenna's radiation pattern [8]. For all results in this section, the operating frequency was 10 GHz unless otherwise stated.

The first demonstration with the metasurface antenna was to generate a broadside beam. The measured farfield pattern for broadside beam generation is shown in Figure 5. Cross sections in azimuth and elevation show the beam pattern and sidelobe level. From the 2D plot, it can be seen that there are no significant grating lobes.

The antenna was also tuned to generate a beam steered in azimuth and elevation. Figure 6a shows the beam steered to 15° in azimuth; Figure 6b shows the beam steered to 15° in elevation. Figure 6c and 6d show beams steered diagonally. Still, no grating lobes appear in the 2D beam patterns, indicating that using feed phase diversity to suppress grating lobes is functioning as intended.

Next, steering performance was more rigorously measured. In azimuth, a beam was steered continuously from

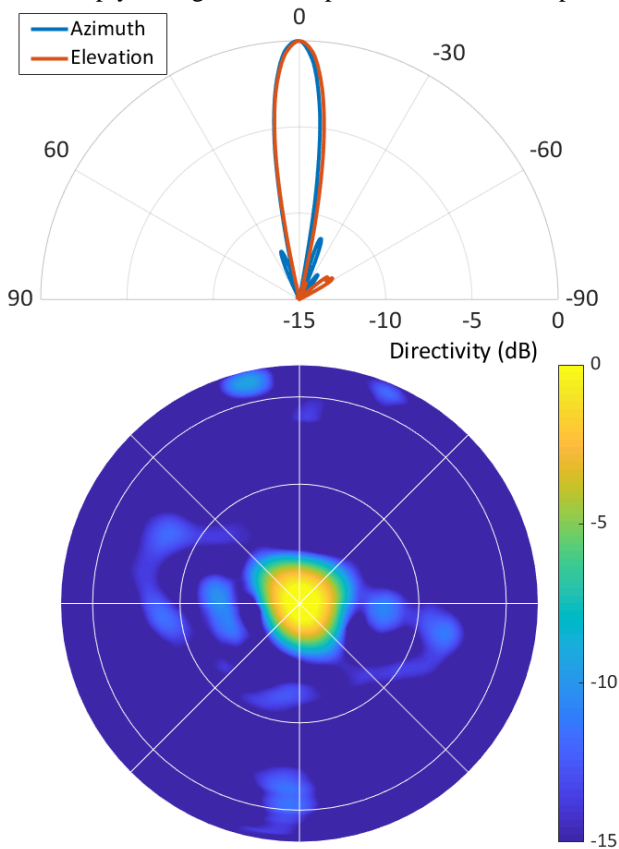

**Figure 5 | Broadside farfield radiation pattern (measured in normalized directivity) of the Nyquist metasurface antenna at 10 GHz. Farfields in all figures are in the u-v plane, where u=cosθcosϕ and v=cosθsinϕ, with gridlines showing θ at 30° and 60° and ϕ from 0° to 360° in 45° increments.**

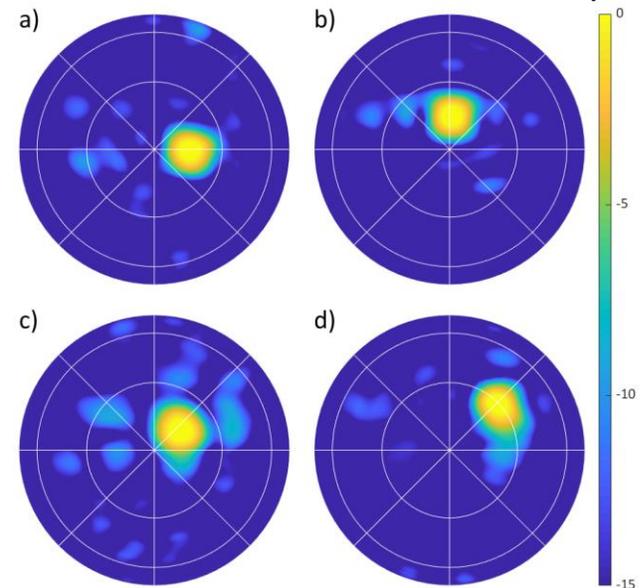

**Figure 6 | Each plot shows the normalized directivity (dB) radiation pattern from the Nyquist metasurface antenna. a) is steered in azimuth to 15°; b) is steered in elevation to 15°. c) and d) show steering in both azimuth and elevation to (10°, 10°) and (30°, 30°), respectively.**

Center for Metamamaterials and Integrated Plasmonics, Department of Electrical and Computer Engineering
Duke University, Durham, North Carolina 27708, USA

-60° to 60°, shown in Figure 7. In elevation, a beam was steered continuously from -75° to 75°, as shown in Figure 8. In both directions, the steering accuracy is shown to the right of the cross section plots. The determined steering ranges are ±50° in azimuth and ±70° in elevation.

To explore the operational bandwidth of the antenna, we generated a broadside beam at various frequencies near 10 GHz. Our target bandwidth of operation was 9.6-10 GHz. Note that performance away from 10 GHz may be degraded due to the feed structure, which was mechanically set for operation at 10 GHz. From 9-11 GHz, a new tuning state was applied and the radiation pattern was measured. Figure 9 shows that the antenna can generate a broadside beam across a large frequency range, from 9.00 to 10.75 GHz.

Though azimuth steering, elevation steering, and operational bandwidth were the primary performance metrics, other metrics were measured. The efficiency of the antenna was measured to be 11% (recorded for broadside beam generation). Polarization isolation was measured to be 30 dB. It should also be noted that all of the tuning states used here were generated by the dipole model described above, without modification to account for fabrication tolerance, mutual coupling, or other effects. Numeric optimization of tuning states could also be used to improve performance but such studies are beyond the scope of this work. The overall antenna performance metrics are summarized below in Table 1.

To explore the capabilities of the antenna further, we generated multiple beams simultaneously. To determine the tuning state for multiple beams, we averaged the $\alpha_{desired}$ for each beam and then applied scaled Euclidean modulation. Still, with this relatively rudimentary effort, two beams could be formed simultaneously in azimuth and in elevation. This experiment yielded the beam patterns shown in Figure 10.

## Discussion

Dynamically reconfigurable antennas play a critical role in many critical technologies, including radar, microwave and imaging, communications, synthetic aperture radar, and many others. Yet, capabilities in many of these fields have been hindered by the high cost and complexity of electronically scanned antenna systems, which have relied predominantly on active components. Metasurface antennas have offered an alternative to traditional ESAs, providing nearly equivalent performance with passive components, avoiding the phase shifters and amplifiers in conventional systems. Where previous systems required dense element spacing and lossy dielectrics, the Nyquist metasurface antenna approach avoids these requirements. The ability to form metasurface antennas using lightweight PCB components unlocks the potential for use in mobile or power-limited environments. Further, Nyquist metasurfaces can be built with hollow waveguides, allowing for future antennas to be highly efficient. The sparse layout presented here scales far better—both in cost and power draw—to extremely large apertures such as those that might be needed in massive MIMO systems, providing a realistic path towards satellite constellations for Earth observation and wireless communications.

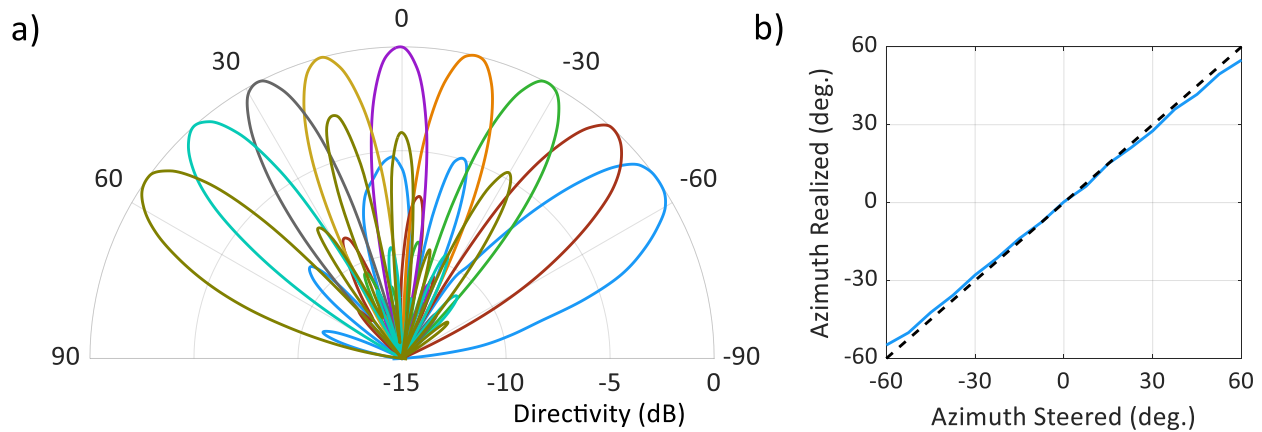

**Figure 7 |** Azimuth steering of the Nyquist metasurface antenna (φ=0°), plotted in normalized directivity. a) shows cross sections steered from -60° to 60° in 15° increments. b) shows the azimuth steering accuracy (with the dashed line showing the goal), indicating that the antenna can accurately steer from -50° to 50°.

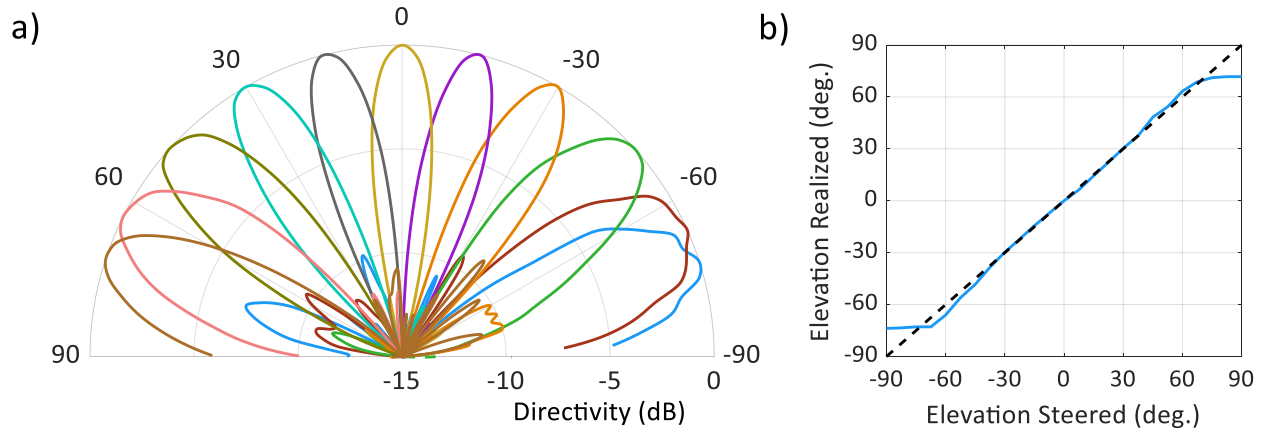

**Figure 8 |** Elevation steering of the Nyquist metasurface antenna (φ=90°), plotted in normalized directivity. a) shows cross sections steered from -75° to 75° in 15° increments. b) shows the elevation steering accuracy (with the dashed line showing the goal), indicating that the antenna can accurately steer from -70° to 70°.

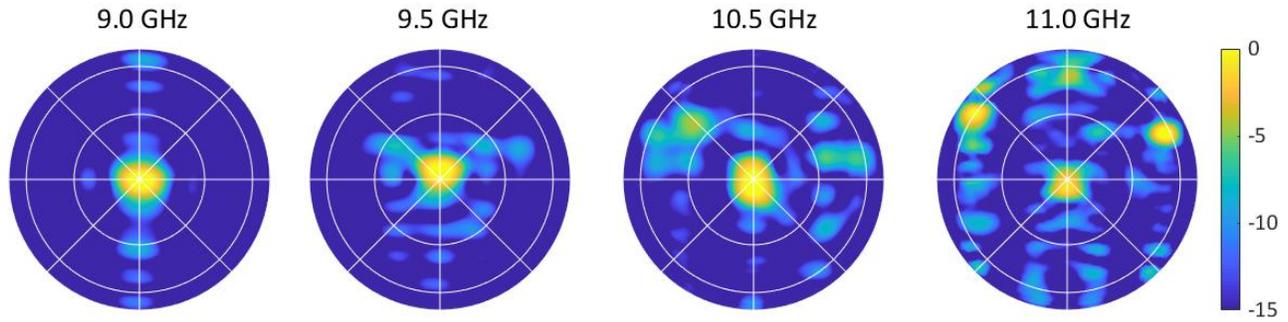

**Figure 9 | Frequency coverage of the Nyquist metasurface antenna, determined by broadside beam formation. For each measurement, a new tuning state has been applied. Plots show normalized directivity (dB).**

**Table 1 | Targeted and realized antenna metrics**

| Metric | Goal | Realized |
| --- | --- | --- |
| Bandwidth | 9.6 – 10 GHz | 9.00 – 10.75 GHz |
| Azimuth steering | +/- 20° | +/- 50° |
| Elevation steering | +/- 20° | +/- 70° |
| Sidelobe level | -13 dB | -12 dB |
| Efficiency | N/A | 11% |
| Gain | N/A | 10.8 dB |
| Polarization | N/A | 30 dB |

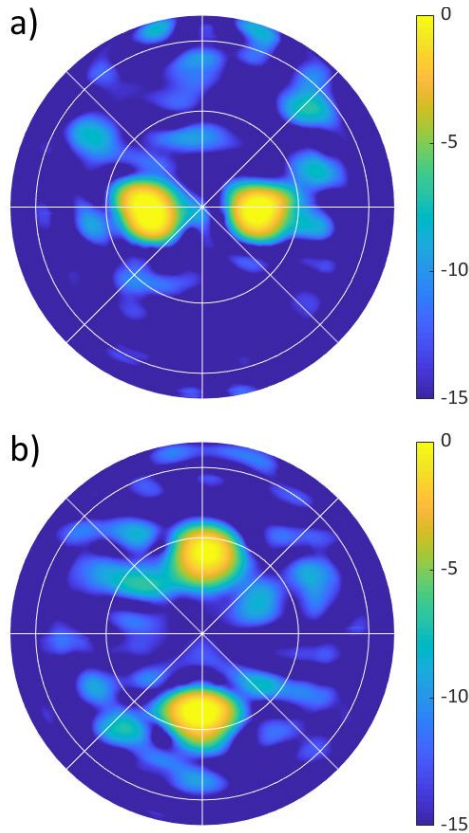

**Figure 10 | Multiple-beam formation with the Nyquist metasurface antenna, measured in normalized directivity. a) shows two beams in azimuth (steered to ±15°); b) shows two beams in elevation (steered to ±25°).**

### Author Contributions
M.B. designed and simulated the antenna with input from T.S. T.S. designed the control circuitry. M.B. and M.F.I. designed the feed structure. M.B. wrote the control software and took measurements, with help from J.N.G. M.B. prepared the manuscript, with input from all other authors.

### Acknowledgements
This work was supported by the Air Force Office of Scientific Research (AFOSR), grant number FA9550-18-1-0187. The authors would like to thank Russell Hannigan, K. Parker Trofatter, and Dr. Daniel Marks.

Center for Metamamaterials and Integrated Plasmonics, Department of Electrical and Computer Engineering
Duke University, Durham, North Carolina 27708, USA

Center for Metamamaterials and Integrated Plasmonics, Department of Electrical and Computer Engineering
Duke University, Durham, North Carolina 27708, USA